\documentclass[a4paper,12pt]{article}
\pdfoutput=1 

\usepackage{jheppub} 

\usepackage[T1]{fontenc} 
\usepackage{physics}
\NewDocumentCommand{\tens}{t_}
{%
	\IfBooleanTF{#1}
	{\tensop}
	{\otimes}%
}
\NewDocumentCommand{\tensop}{m}
{%
	\mathbin{\mathop{\otimes}\displaylimits_{#1}}%
}
\usepackage{graphicx}
\usepackage{wrapfig}
\usepackage{amsmath}
\usepackage{amsfonts}
\usepackage{amssymb}
\usepackage{quotes}
\usepackage{subcaption}
\usepackage{calligra}
\newcommand{\rc}{\textit{\huge\calligra r}}
\usepackage{changepage} 

\usepackage{afterpage}

\usepackage{placeins}
\newcommand{\z}{\Bar{z}}
\newcommand{\paren}[1]{\left(#1\right)}
\newcommand{\m}{\Tilde{m}}

\newcommand{\T}[1]{\tilde{#1}}

\title{\textbf{\boldmath One-point holographic correlator in the expanding universe}}

\author{Souvik Paul,}
\author[1]{Gopinath Guin,\note{Corresponding author.}}
\author{Sunandan Gangopadhyay}

\affiliation[a]{\textit{Department of Astrophysics and High Energy Physics\\}
	\textit{ S.N.~Bose National Centre for Basic Sciences,\\}
	\textit{JD Block, Sector-III, Salt Lake, Kolkata 700106, India}}

\emailAdd{souvik.paul@bose.res.in}
\emailAdd{gopinath.guin@bose.res.in}
\emailAdd{sunandan.gangopadhyay@bose.res.in}

\abstract{\noindent In this article, we have calculated the time-dependent thermal one-point function of massive operators within an expanding universe. We employ the Randall-Sundrum II braneworld model combined with a $p$-brane gas in the bulk, enabling us to represent various matter-dominated cosmological scenarios localised on the brane. Applying the geodesic formula introduced by Grinberg and Maldacena in \cite{Grinberg:2020fdj}, we have calculated the thermal one-point functions of massive operators within the universe. The time-dependent one-point functions for different matter-dominated universes, both single-component and multi-component, are derived from the brane's evolving radial position. The time-dependent positions of the branes have been obtained using the second Israel junction condition. Additionally, we have analyzed the early and late-time behaviors of the thermal one-point function across various matter-dominated universes.}
\begin{document}
	\maketitle
	\flushbottom
	\section{Introduction}
\begin{figure}
    \centering
    \includegraphics[width=0.8\linewidth]{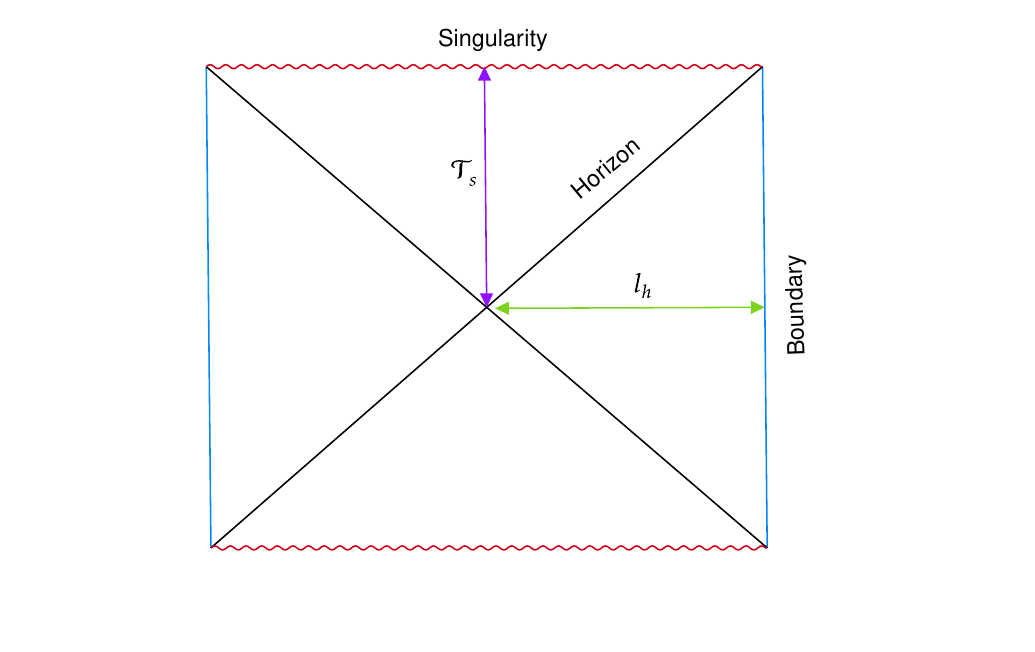}
    \caption{Penrose diagram of an eternal AdS-Schwarzschild black hole. Here, the blue lines indicate the AdS boundary where the conformal field theory lives. The wiggly red lines denote the black hole singularity. Here $l_h$ is the regularised geodesic length from the boundary to the horizon and $\mathcal{T}_s$ denotes the time-like distance between the black hole horiozn to singularity.}
    \label{fig:opf penrose diagram}
\end{figure}

    The AdS/CFT correspondence \cite{Maldacena:1997re,Gubser:1998bc,Witten:1998qj,Aharony:1999ti,Natsuume:2014sfa,nuastase2015introduction,Hubeny:2014bla} acts as a tool to study strongly interacting theories with quantum effects from a weakly interacting gravity theory in one extra dimension. The extra bulk radial direction refers to the direction of the renormalisation group (RG) flow of the dual quantum field theory (QFT). For the past few years, this correspondence has been used to study the correlations in strongly interacting systems.  For a large-$N$ QFT which has a bulk description, the two-point correlation function of two boundary operators can be obtained holographically from the following path integral \cite{Balasubramanian:1999zv,Banks:1998dd,Louko:2000tp,Susskind:1998dq}
\begin{equation}
    \langle\mathcal{O}(t,x)\mathcal{O}(t,y)\rangle=\int \mathcal{D}\mathcal{P}~e^{-\Delta \mathcal{L}(\mathcal{P})}
\end{equation}
where $\mathcal{L}(\mathcal{P})$ is the path in the bulk geometry which connects the boundary operators and $\Delta$ is the conformal dimension of the boundary operators. Saddle point approximation of the above path integral around the classical turning point gives
\begin{equation}
    \langle\mathcal{O}(t,x)\mathcal{O}(t,y)\rangle\sim e^{-\Delta L(t,x;t,y)}
\end{equation}
where $L(t,x;t,y)$ is the geodesic length between the boundary operators $\mathcal{O}(t,x)$ and $\mathcal{O}(t,y)$. It should be mentioned that the above expression for the two-point correlation function is only valid for operators with large conformal dimensions. Whether the above two-point correlator is a thermal one or not, depends upon the fact whether the bulk geometry contains a black hole or not. \\
On the other hand the interior physics of the black hole have continued to be a problem of interest in quantum gravity. In \cite{Grinberg:2020fdj}, it was shown that the thermal one-point function of the boundary operators even contains some information about the physics behind the black hole event horizon. Grinberg and Maldacena considered a model used in \cite{Myers:2016wsu} where a coupling between the bulk scalar field and the squared Weyl tensor is included. A minimally coupled field with a quadratic action results in a one-point function that is zero. However, a non-zero value might arise due to higher-derivative corrections to the action. On a black hole background, this gravitational coupling acts as a source term for the field, thereby resulting in a non-zero one-point function. Using the intuition from these exactly solvable models \cite{Grinberg:2020fdj}, a saddle point approximation was developed for large masses. In the Wentzel–Kramers–Brillouin (WKB) approximation \cite{wentzel1926verallgemeinerung,kramers1926wellenmechanik,brillouin1926mecanique}, the saddle point occurs at complex radial positions, and its location, along with the contour choice, is justified through detailed arguments. Their study reveals that the one-point function of an AdS$_5$ black hole with inner horizon exponentiates to the following form \cite{Grinberg:2020fdj}
\begin{equation}
    \langle\mathcal{O}\rangle\sim (\text{Powers of }m)e^{-im \mathcal{T}_s-ml_h}~~~~\text{for}~Im(m)<0
\end{equation}
where $m$ is the mass of the bulk scalar field, $\mathcal{T}_s$ is the proper time for a particle to reach the singularity from the horizon, and $l_h$ is the regularized proper length from the boundary to the horizon (see Fig.\eqref{fig:opf penrose diagram} for visual understanding). It should be mentioned that the time to the singularity arises from a “phase” in the one-point function. This formula allows one to compute the thermal one-point function of the boundary CFT holographically. Several interesting studies regarding the one and two-point correlation functions for various bulk spacetimes can be found in \cite{Saha:2025hap,Fischler:2012ca,Krishna:2021fus,Keranen:2016ija,Rodriguez-Gomez:2021pfh,Grinberg:2020uht,Park:2024pkt,Kim:2023fbr,Park:2022mxj,Berenstein:2022nlj,David:2023uya,Georgiou:2022ekc,Dodelson:2022yvn,Park:2022abi,Dey:2026ckh}. For anisotropic Kasner universes, people have computed different correlation functions, which can be found in \cite{Caceres:2023zhl,Frenkel:2020ysx,Caputa:2021pad,Banerjee:2015fua}.
It is obvious to ask about the correlations between different massive operators in our expanding universe. Recently, in \cite{Park:2024omh}, the authors have studied about the two-point correlation functions for different matter-dominated expanding universes. They have used the back reaction of the $p$-brane gas in the bulk to introduce various matter sources in the universe, which is situated on the brane. The geodesic formula had been used to compute the two-point correlation function of the expanding universe in a holographic manner. The time dependence in the expression of the two-point correlation function is included through the time-dependent brane position, which was obtained through the second Israel junction condition. Although their study does not shed any light on the thermal one-point function of the expanding universe in the presence of various matter sources. In this paper, we have computed the thermal one-point function for massive operators in the expanding universe. We have used the geodesic formula proposed by Grinberg and Maldacena \cite{Grinberg:2020fdj} to compute the one-point function of our universe. In the braneworld model, we have inserted various matter sources through different configurations of $p$-brane gas in the bulk. Then, the time-dependent one-point function is calculated by using the time-dependent radial position of the brane. The time-dependent brane positions are computed through the second Israel junction condition for various matter-dominated universes. We have also graphically expressed the time evolution of one-point functions of massive operators in the expanding universe.\\
The rest of the paper is organised as follows. In section \eqref{sec 2}, we have given a brief review of the Randall–Sundrum II braneworld model and derived the second Israel junction condition. The second Israel junction condition eventually gives us the time dynamics of the brane's radial position. Section \eqref{sec 3} is dedicated to computing the time dependence of the brane position through $p$-brane gas. Here we have shown how different $p$-brane gas configurations in the bulk back react on the brane and generate various single and multi-component universes. In section \eqref{sec 4}, we have given a quick recap on the derivation of the geodesic formula for the thermal one-point function of massive boundary operators. Then, by considering a black brane geometry in the bulk, we have calculated the one-point functions for massive operators on the brane. We have also calculated the early and late time behavior of the one-point function for various matter-dominated universes for both single and coexisting matter components. Finally, we conclude in section \eqref{sec 5}.
\section{Randall–Sundrum braneworld model and Israel junction condition}\label{sec 2}
The braneworld model in cosmology has become a very interesting way to study the universe and its properties. In this study, we have taken the RS-II braneworld, where the FLRW ((Friedmann–Lemaître–Robertson–Walker \cite{Friedman:1922kd,Friedmann:1924bb,Lemaitre:1931zza,Lemaitre:1933gd,Robertson:1935jpx,Robertson:1935zz,Walker:1937qxv}) universe, being a four ($3+1$)-dimensional object, is situated on the four-dimensional brane. The brane is embedded in a five-dimensional manifold. The extra one dimension orthogonal to the brane is the radial direction. The expansion of the universe can be thought of as the motion of the brane in the radial direction. As the brane is four-dimensional and the manifold is five-dimensional, the brane slices the manifold into two regions at a particular value of the radial position of the brane. The bulk spacetime on both sides of the brane (say $\mathcal{M_+}$ and $\mathcal{M_-}$) must be joined smoothly, by an appropriate (Israel) junction condition to keep the manifold intact. There are two Israel junction conditions, the first one says that the metric on the two sides of the brane must match on the brane to have a unique metric of the brane. The second Israel junction condition \cite{Israel:1966rt} deals with the existence of any source terms on the brane. We will show this condition mathematically in the following section.
To start with, we take the Einstein-Hilbert action with the GHY boundary term
\begin{equation}
    S=\frac{1}{16\pi G_5}\int_\mathcal{M} d^5x\sqrt{-g}\left(R-2\Lambda\right)+\frac{1}{8\pi G_4}\int_{\partial\mathcal{M}}d^4x\sqrt{-h}\mathcal{K}
\end{equation}
where $g$ and $h$ are the determinants of the total and boundary metric. $R,\Lambda$ are the Ricci scalar and cosmological constant, respectively. $\mathcal{K}$ is the trace of the extrinsic curvature of the brane. Varying the above action, after a little lengthy calculation, one gets \footnote{The reader is encouraged to refer to \cite{10.1093/ptep/ptag115,Paul:2025gpk,Chamblin:1999ya} for a comprehensive account of the calculations.}
\begin{equation}
    \delta S=-\frac{1}{16\pi G_4}\int_{\partial\mathcal{M}}d^4 x\sqrt{-h}\Big[\mathcal{K}^{\mu\nu}-\mathcal{K} h^{\mu\nu}\Big]\delta    h_{\mu\nu}~.
\end{equation}
This now straightforwardly gives the energy-momentum tensor for the bulk regions $\mathcal{M_\pm}$
\begin{equation}
    T^{(\pm)}_{\mu\nu}=-\frac{1}{16\pi G_4}\left(\mathcal{K}_{\mu\nu}^{(\pm)}-\mathcal{K}^{(\pm)}h_{\mu\nu}\right)~.
\end{equation}
Considering the $Z_2$ symmetry of the brane, we have $\mathcal{K}_{\mu\nu}^{(+)}=-\mathcal{K}_{\mu\nu}^{(-)}$. By renaming $\mathcal{K}_{\mu\nu}^{(+)}=\mathcal{K}_{\mu\nu}$, we can write
\begin{equation}\label{energy momentum tensor}
    T^{(\pm)}_{\mu\nu}=\pm\frac{1}{16\pi G_4}\left(\mathcal{K}_{\mu\nu}-\mathcal{K}h_{\mu\nu}\right)~.
\end{equation}
The above expression gives the energy-momentum tensor on both sides of the brane. So the energy-momentum tensor of the brane is the difference between the two. Mathematically,
\begin{equation}\label{energy momentum tensor of the brane}
    T^{brane}_{\mu\nu}=T_{\mu\nu}^{(+)}-T_{\mu\nu}^{(-)}~.
\end{equation}
On the other hand, the action of the brane (in the ground state) is given as \cite{Chamblin:1999ya}
\begin{equation}
    S_{brane}=-\frac{\sigma}{4\pi G_4}\int_{\partial\mathcal{M}}d^4x \sqrt{-h}
\end{equation}
where $\frac{\sigma}{4\pi G_4}$ is the brane tension. Varying this action, one gets the energy-momentum tensor
of the brane to be
\begin{equation}\label{einstein tensor}
    T^{brane}_{\mu\nu}=\frac{\sigma}{8\pi G_4}h_{\mu\nu}~.
\end{equation}
The above energy-momentum tensor(eq.\eqref{einstein tensor}) must be equal to the energy momentum tensor in eq.\eqref{energy momentum tensor of the brane}, which can be written as using the form in eq.\eqref{energy momentum tensor} as
\begin{equation}
   \mathcal{K}_{\mu\nu}-\mathcal{K}h_{\mu\nu}=\sigma h_{\mu\nu}~.
\end{equation}
The above equation can be simplified further by taking the trace on both sides, which gives $\mathcal{K}=-\frac{4\sigma}{3}$, and therefore finally we arrive at
\begin{equation}\label{extrinsic}
    \mathcal{K}_{\mu\nu}=-\frac{\sigma}{3}h_{\mu\nu}~.
\end{equation}
Again, the form of the extrinsic curvature of a geometry is defined as 
\begin{equation}\label{extrinsic curvature}
    \mathcal{K}_{\mu\nu}=h^\alpha_\mu h^\beta_\nu\nabla_\alpha n_\beta
\end{equation}
where $n$ is the unit normal vector of the associated geometry. Now to find out the form of the unit normal vector, we must know the exact form of the metric. Here we proceed by taking a simple metric which has translational and rotational symmetries and has the following form
\begin{equation}\label{metric}
ds^2=-A(r) dt^2+B(r)dr^2+C(r)\delta_{ij}dx^idx^j    
\end{equation}
here $i,j=1,2,3$. With this metric, the normal vector has the following form
\begin{equation}
    n_{\alpha}=\frac{\sqrt{A(r)B(r)}}{\sqrt{A(r)-B(r)\dot r^2}}\left(\dot r,-1,0,0,0\right)
\end{equation}
where the dot over the $r$ implies the derivative of $r$ with respect to $t$. If we put the above expression of the normal vector into the expression of the extrinsic curvature in eq.\eqref{extrinsic curvature}, we get the spatial components of it as
\begin{equation}
    \mathcal{K}_{ij}=-\frac{1}{2\sqrt{A(r)-B(r)\dot r^2}}\sqrt{\frac{A(r)}{B(r)}}\frac{C'(r)}{C(r)}h_{ij}
\end{equation}
where the prime $\prime$ denotes derivative with respect to $r$.
Using the form of the extrinsic curvature from eq.\eqref{extrinsic}, the above equation can be recast to
\begin{equation}
    \frac{C'}{C}=\frac{\sigma}{3}\sqrt{\frac{B(r)}{A(r)}}\sqrt{A(r)-B(r)\dot r^2}~.
\end{equation}
This immediately gives
\begin{equation}\label{r dot}
    \dot r^2=\frac{A}{B}\left(1-\frac{9}{\sigma^2G}\frac{H'^2}{H^2}\right)~.
\end{equation}
Now to do calculations, we have to project the metric into the form of the FLRW metric. To do so, we take the following parameterisation
\begin{equation}\label{transormation equation}
    -d\tau^2=-A(r)dt^2+B(r)dr^2~.
\end{equation}
With this, we can write the metric (eq.\eqref{metric}) in the following form
\begin{equation}
    ds^2=-d\tau^2+a(\tau)\delta_{ij}dx^ide^j 
\end{equation}
where $H(r)$ has been transformed according to the coordinate transformation and identified as $a(\tau)$. 
Using the transformation eq.\eqref{transormation equation}, one can recast eq.\eqref{r dot} as follows
\begin{equation}\label{Isreal JC}
    \paren{\frac{dr}{d\tau}}^2=\frac{\sigma^2}{9}\frac{H^2(r)}{H'^2(r)}-\frac{1}{G(r)}~.
\end{equation}
With some known form of $H(r)$ and $G(r)$, the above equation, after integration, gives the motion of the brane up to some integration constant. Using the initial condition, one again fixes the integration constant.
In the next part, we will see that the choice of some $G(r)$ and $H(r)$ can introduce radiation, matter or other exotic types of elements either in single-component form or multi-component form in the universe.
\section{Time dependence of the brane position through p-brane gas}\label{sec 3}    
Until now, we have considered the general form of the metric with time translational symmetry. Now, to have different types of source terms, we have considered $p$-brane string gas in the bulk. The different types of p-brane gas backreact with the brane and introduce different types of elements of the universe. The Einstein-Hilbert action with the p-brane gas geometry has the form
\begin{equation}
    S=\frac{1}{16\pi G_5}\int d^5x \sqrt{-g}\paren{R-2\Lambda}+ T_pN_p\int d^{p+1}\xi \sqrt{-h}h^{\alpha\beta}\partial_\alpha x^\mu \partial_\beta x^\nu g_{\mu\nu}
\end{equation}
where $T_p$ and $N_p$ are the tension and number of p-branes, respectively. $x^\mu$'s are the bulk coordinates and $\xi^\mu$'s are the coordinates of the brane. $h_{\alpha\beta}$ is the induced metric on the brane. \\
Varying the above action, we get the Einstein equation as
\begin{equation}
  R_{\mu\nu}-\frac{1}{2}Rg_{\mu\nu}+\Lambda g_{\mu\nu}=16\pi G_5T_{\mu\nu}  
\end{equation}
where $R_{\mu\nu}$ and $R$ are the Ricci tensor and its trace, respectively. In the above equation, the energy-momentum tensor has the form 
\begin{equation}
    T_{\mu\nu}=-T_pn_P\Bigg\{g_{tt},\frac{p-1}{3}g_{11},\frac{p-1}{3}g_{22},\frac{p-1}{3}g_{33},g_{rr}\Bigg\}
\end{equation}
where $n_p$ is the number density. The most general spacetime for a negative cosmological constant is an AdS black brane, for which we start with the metric ansatz 
\begin{equation}
    ds^2=\frac{r^2}{\tilde R^2}\Big(-f(r)dt^2+\delta_{ij}dx^idx^j\Big)+\frac{R^2}{r^2f(r)}dr^2~,
\end{equation}
where the indices $i,j$ have values $1,2,3$ and $f(r)$ is the lapse function for the black brane geometry. Using the energy-momentum tensor and the above ansatz of the black brane geometry, we can obtain the Einstein field equations, which are differential equations of the lapse function. Solving for the $tt$-component, we have the lapse function 
\begin{equation}\label{bb lapse function}
    f(r)=1-\frac{\rho_p}{r^{4-p}}-\frac{C}{r^4}~,
\end{equation}
where $\rho_p=8\pi G c_p n_p T_p$,
$\rho_p,C$ being the energy density of the string gas and an arbitrary integration constant, $c_p$ determined by the value $p$. In the absence of string gas, one can easily see that the last term is like the Schwarzschild term and, in fact, $C$ can be identified with the ADM mass of the black hole \cite{Arnowitt:1959ah}. For this reason, we rename $C$ as $\m$. The second term arises entirely from the backreaction of the string gas in the bulk.\\
In earlier works in this direction \cite{Park:2021wep,Paul:2025gpk}, it was shown that in the absence of a Schwarzschild-like term, different kinds of string gas can produce dark and exotic matter, and in the absence of string gas, there is radiation. The existence of both terms gives radiation and either dark matter or exotic matter-like behavior. 
\subsection{Time dependence of the brane in a single-component universe}
Now we will proceed further to compute the time-dependent brane positions for various single and multi-component universes on the brane. To do this, we will use the first-order differential equation in eq.\eqref{Isreal JC}, governing the brane dynamics along with different combinations of the black brane lapse function from eq.\eqref{bb lapse function}.
\subsubsection{Radiation dominated universe} 
In order to find the time dependence of the brane position, we use the lapse function eq.\eqref{bb lapse function} of the brane and use the Israel junction condition eq.\eqref{Isreal JC}. To get the time dependence for the radiation-dominated universe, proceed by taking the lapse function in the absence of the string gas, which implies 
\begin{equation}
    \paren{\frac{dr}{d\tau}}^2=\paren{\frac{\sigma^2}{36}-\frac{1}{\tilde R^2}}r^2+\frac{\m}{r^2R^2}~.
\end{equation}
Now, to get the nontrivial contribution, we set the brane tension at the critical value $\sigma=\sigma_c$ for which we write 
\begin{equation}
    \Big({\dfrac{dr}{d\tau}}\Big)^2=\frac{\m}{r^2R^2}~.
\end{equation}
Integrating the above equation, we get the time-dependent brane position 
\begin{equation}
    r(\tau)=\frac{\sqrt{2}\m ^{1/4}}{\sqrt{R}}\tau^{1/2}+r_i
\end{equation}
$r_i$ being the brane position at $t=t_i=0$ that is the initial brane position. So we see the brane position scales with the time as $\tau^{1/2}$.
Now, defining the inverse radial coordinate $\z(\tau)\equiv1/r(\tau)$ we get the relation of the brane position in inverse coordinate with respect to time \cite{Park:2020jio,Paul:2025gpk}
\begin{equation}\label{z}
    \z(\tau)=\frac{z_hz_i}{z_h+\sqrt{2}z_i\tau}~,
\end{equation}
where we have defined $z_i=1/r_i$ and $z_h$ is the position of the horizon of the black brane in inverse coordinate and has the equivalent definition $z_h=\m^{-1/4}$. 
Now for early time $z_h>z_i\tau$, which gives approximately
\begin{equation}
    \z(\tau)\simeq z_i\paren{1-\frac{\sqrt{2}z_i}{z_h}\tau}~.
\end{equation}
Also, for the late time we have $\tau\rightarrow\infty$, and using eq.\eqref{z} we write the brane position in inverse radial coordinate in terms of the time as
\begin{equation}
    \z(\tau)\simeq\frac{z_h}{\sqrt{2}\tau}~.
\end{equation}
Clearly, in the late time, the brane position in the inverse coordinate scales inversely with time.
\subsubsection{Matter dominated universe}
To get the brane dynamics with time, we set $\m=0,p=1$ and take critical brane tension to have 
\begin{equation}
    \Big({\dfrac{dr}{d\tau}}\Big)^2=\frac{\rc}{rR^2}~.
\end{equation}
After integrating the above equation, we get the time-dependent brane position for a universe dominated by dark matter only as
\begin{equation}
    r(\tau)=\paren{\frac{3}{2}}^{2/3} \paren{\frac{\rc}{R^2}}^{1/3}\tau^{2/3}+r_i ~.
\end{equation}
Clearly, the brane scales with time as $\tau^{2/3}$. This scaling is similar to the scale factor in a matter-dominated universe for standard cosmology. Now, similar to the previous case of a radiation-dominated universe, we define the inverse radial coordinate $\z(\tau)\equiv1/r(\tau)$ and also defining $z_i=r_i^{-1}$; we have \cite{Park:2020jio,Paul:2025gpk}
\begin{equation}\label{z_matter}
    \z(\tau)=\frac{z_i}{1+\left(\frac{3}{2}\right)^{2/3}\rc^{1/3} z_i \tau^{2/3}}~.
\end{equation}
We now find out the leading order behavior in the early and late time scenarios. In the early time, the second term in the denominator is less than one, implying 
\begin{equation}
    \z(\tau)\simeq z_i\paren{1-\left(\frac{3}{2}\right)^{2/3}\rc^{1/3} z_i \tau^{2/3}}~.
\end{equation}
For the late time, we have $\tau \rightarrow\infty$, using which we can write (from eq.\eqref{z_matter})
\begin{equation}
    \z(\tau) \simeq \frac{\paren{\frac{2}{3}}^{2/3}}{\rc^{1/3}\tau^{2/3}}~.
\end{equation}
Clearly, in the late time, the brane position in the inverse radial coordinate scales as $\tau^{-2/3}$ in the late time.
\subsubsection{Exotic matter dominated universe}
If we take the Schwarzschild part of the lapse function equals zero in eq.\eqref{bb lapse function} and take $p=2$ with critical brane tension, we get
\begin{equation}
\Big({\dfrac{dr}{d\tau}}\Big)^2=\frac{\delta}{R^2}~.
\end{equation}
Integrating this we obtaine 
\begin{equation}
    r(\tau)=\frac{\sqrt{\delta}}{R}\tau+r_i ~.
\end{equation}
So, for exotic matter the brane grows linearly with time. 
Similar to the previous two cases, here we also takes the inverse radial coordinate $\z(\tau)\equiv r(\tau)^{-1}$ and $z_i=r_i^{-1}$. With these definitions, we can write from the previous equation \cite{Park:2020jio,Paul:2025gpk}
\begin{equation}
    \z(\tau)=\frac{z_i}{1+\sqrt{\T\delta}\tau z_i}~,
\end{equation}
which in the early times gives 
\begin{equation}
    \z(\tau)=z_i\paren{1-\sqrt{\delta}\tau z_i}~.
\end{equation} 
Similarly in the late time ($\tau \rightarrow \infty$), we get
\begin{equation}
    \z(\tau)=\frac{1}{\sqrt{\delta}\tau}~.
\end{equation} 
Hence, the brane position in inverse radial coordinate scales inversely with the cosmological time $\tau$. \\
So for all the single-component universes, we have calculated the brane position in inverse coordinates in terms of time. We have figured out the leading-order behavior both in the early and late time scenarios. These results will be necessary to figure out the time-dependent thermal one-point functions in the later sections.
\subsection{Time dependence of the brane in a multi-component universe}
For multi-component universes, we have to use the full black brane lapse function in eq.\eqref{bb lapse function} along with the Israel junction condition to obtain the time-dependent brane positions. It is already discussed that the black brane lapse function always has a Schwarzschild-like term, which gives radiation on the brane and one extra term that comes from the $p$-brane gas. This extra term gives rise to matter or exotic matter on the brane. Due to this, we will discuss about universes with coexisting radiation-matter and radiation-exotic matter.
\subsubsection{Radiation and matter-dominated universe}
In a more general case, we can take a non-vanishing $\m$ with the string gas to have a multi-component universe. For such a case, a universe with $p=1$ we have
\begin{equation}\label{dr_dt}
    \paren{\frac{dr}{d\tau}}^2=\frac{m}{r^2R^2}+\frac{\rc}{R^2r}~.
\end{equation}
The above equation, after integration gives 
\begin{equation}
    \paren{\frac{r^2}{2}-\frac{r^2_i}{2}}-\paren{\frac{r^3}{6r_t}-\frac{r_i^3}{6r_t}}=\tau\sqrt{m}~.
\end{equation}
where $r_t=m/\rc$, is the matter-radiation equality moment. 
The above equation, upon solving perturbatively for the early time, we get
\begin{equation}
    r\simeq r_i+\frac{\sqrt{m}\;\tau}{r_i \tilde R}+\frac{\sqrt{m}\;\tau}{2\tilde Rr_t}~.
\end{equation}
Projecting this into the inverse radial coordinate, similar to the single-component universe, we define $z(\tau)\equiv1/r(\tau)$ along with $z_i=1/r_i$ and $z_t=1/r_t$. This gives \cite{Paul:2025gpk}
\begin{equation}\label{early_brane_rad+matt}
    \z\simeq z_i\paren{1-\sqrt{m}\;\tau\paren{z_i^2+\frac{z_iz_t}{2}}}~.
\end{equation}
The above equation shows how the brane position in the inverse radial coordinate scales with time.\\
In the case of late time($\tau\rightarrow\infty$) integrating perturbatively the eq.\eqref{dr_dt} gives 
\begin{equation}
    r\simeq\paren{\frac{3}{2}}^{2/3} \rc^{1/3}\tau^{2/3}\Bigg(1-\paren{\frac{2}{3}}^{2/3} \frac{r_t}{\rc^{1/3}}\tau^{-2/3}+\frac{2}{3\sqrt{m}}\frac{1}{\tau}\paren{\frac{2r_t^2}{3}-\frac{r_i^2}{2}+\frac{r_i^3}{6r_t}}\Bigg)~.
\end{equation}
In the inverse radial coordinate which look like \cite{Paul:2025gpk}
\begin{equation}\label{late_brane_rad+matt}
        \z \simeq \paren{\frac{2}{3}}^{2/3}\rc^{-1/3}\tau^{-2/3}\Bigg(1+\paren{\frac{2}{3}}^{2/3} \frac{r_t}{\rc^{1/3}}\tau^{-2/3}-\frac{2\sqrt{z_t}}{3\sqrt{\rc}}\frac{1}{\tau}\paren{\frac{2}{3z_t^2}-\frac{1}{2z_i^2}+\frac{z_t}{6z_i^3}}\Bigg)~.
\end{equation}
This is the brane dynamics in the late-time scenario.
It is worth mentioning that at a late time, one can not demand $\tau\rightarrow0$, hence $\z \rightarrow z_i$.
\subsubsection{Radiation and exotic matter dominated universe}
Similar to the radiation-matter case, if you set $p=2$ in the lapse function, we get for the early time
\begin{equation}\label{int eq rad+ex}
    \paren{\frac{dr}{d\tau}}^2=\frac{m}{r^2}+\chi = \frac{m}{r^2}\left(1+\frac{r^2}{r_t^2}\right)~.
\end{equation}
In the early time, as $r\ll r_t$, $r_t$ being the radiation exotic matter equality point, we have perturbatively (up to order of $\tau$)
\begin{equation}
r=r_i\Bigg[1+\sqrt{m}\paren{\frac{1}{r_i^2}-\frac{1}{2\T r_t^2}}\tau\Bigg]~.
\end{equation}
In the inverse radial coordinate this perturbatively gives \cite{Paul:2025gpk}
\begin{equation}\label{early brane rad+ex}
    \z=z_i\Bigg[1-\sqrt{m}\paren{{z_i^2}-\frac{ z_t^2}{2}}\tau\Bigg]~.
\end{equation}
Again in the late time, after integrating perturbatively eq.\eqref{int eq rad+ex}, we get \cite{Paul:2025gpk}
\begin{equation}\label{late brane rad+ex}
    \z\simeq\frac{1}{\sqrt{\chi}\tau}\Bigg[1-\frac{1}{\sqrt{m}}\paren{\frac{3}{8 z_t^2}-\frac{1}{2z_i^2}}\frac{1}{\tau}-\frac{1}{2\chi z_t^2 \tau^2}\Bigg]~.
\end{equation}
Clearly, in the late time for large $\tau$, the exotic matter controls the dynamics of the brane as it should be according to the thermal history of the universe \cite{WMAP:2010qai,WMAP:2010sfg,Planck:2014loa,Planck:2018vyg}, as exotic matter mimics the contribution of spatial curvature in the Friedmann equation. 
\section{Holographic calculation of cosmological one-point function}\label{sec 4}
In this section, we compute the cosmological one-point functions for different matter-dominated universes. We have discussed the holographic calculation of the one-point function for both single- and multi-component universes. In \cite{Grinberg:2020fdj}, the authors proposed a formula to compute the thermal one-point function using the geodesic approximation. It is well known that a minimally coupled bulk field has a quadratic action, which leads to a vanishing one-point function. Although one can obtain a non-zero value of the one-point function if there are higher derivative couplings in the action. Some examples of such kind of coupling are the coupling between the bulk scalar field and the Weyl tensor square or the Gauss-Bonnet tensor. This gravitational coupling leads to a source term for the field and, therefore, a non-vanishing one-point function. In an asymptotically AdS$_{d+1}$ background, a minimally coupled scalar field $\phi$ of mass $m$ is dual to a boundary operator $\mathcal{O}$ in one lower dimension. The boundary operator has a conformal dimension \cite{Gubser:1998bc,Witten:1998qj}
\begin{equation}
    \Delta=\frac{d}{2}\pm\sqrt{\frac{d^2}{4}+m^2 \T R^2}
\end{equation}
where $\tilde R$ is the $AdS$ radius. For bulk AdS$_5$ spacetime $d=4$ and the above expression for the conformal dimension $\Delta$ becomes
\begin{equation}
    \Delta=2\pm\sqrt{4+m^2 \T R^2}~.
\end{equation}
From the above expression, it is evident that for the conformal dimension $\Delta$ to be real and positive, $m^2$ must satisfy
\begin{equation}
    m^2\geq -\frac{4}{\T R^2}~.
\end{equation}
The above inequality is the well-known Breitenlohner-Freedman (BF) bound \cite{Breitenlohner:1982bm}. This bound indicates that, despite the bulk scalar field having a negative mass, it remains stable within the AdS spacetime \footnote{This bound is also very useful in the study of holographic superconductors \cite{Hartnoll:2008kx,Hartnoll:2008vx,Herzog:2009xv,Horowitz:2010gk,Li:2011xja,Gangopadhyay:2012am,Paul:2025apr}}. This bound on the conformal dimension will be later used while graphically representing the time dynamics of the thermal one-point functions of various single and multi-component universes in the braneworld cosmological model.\\
In \cite{Grinberg:2020fdj}, the low-energy effective action containing a single massive field and the simplest higher-derivative coupling to the gravitational field is considered, which reads
\begin{equation}
    S=\frac{1}{16\pi G_{d+1}}\int d^{d+1}x \sqrt{-g}\Bigg(\frac{1}{2}\nabla_{\mu}\phi \nabla^{\mu}\phi+\frac{m^2}{2}\phi^2 +\alpha\phi W_{\mu\nu\rho\sigma}W^{\mu\nu\rho\sigma}\Bigg)~
\end{equation}
where $\alpha$ is the coupling parameter and the Weyl tensor and the Weyl tensor square for a $(d+1)$-dimensional spacetime is given by \cite{weyl1918reine,weinberg1973gravitation}
\begin{align}
    W_{\mu\nu\rho\sigma}&=R_{\mu\nu\rho\sigma}+\frac{1}{d-1}\Big(g_{\mu\rho}R_{\nu\sigma}-g_{\nu\rho}R_{\mu\sigma}+g_{\nu\sigma}R_{\mu\rho}-g_{\mu\sigma}R_{\nu\rho}\Big)+\frac{R}{d(d-1)}\Big(g_{\mu\rho}g_{\nu\sigma}-g_{\mu\sigma}g_{\nu\rho}\Big)\\
    W^2&=W_{\mu\nu\rho\sigma}W^{\mu\nu\rho\sigma}=R_{\mu\nu\rho\sigma}R^{\mu\nu\rho\sigma}-\frac{4}{d-1}R_{\mu\nu}R^{\mu\nu}+\frac{2}{d(d-1)}R~.
\end{align}
Here $R_{\mu\nu\rho\sigma}$, $R_{\mu\nu}$ and $R$ are the Riemann tensor, Ricci tensor and Ricci scalar, respectively. More details on the causality-based bounds on the coupling parameter $\alpha$ can be found in \cite{Cordova:2017zej,Meltzer:2017rtf}. As it is a higher derivative coupling, it is suppressed in the strong t’Hooft coupling limit. From dimensional analysis, it was shown that 
\begin{equation}
    \alpha\sim l_{s}^2\sim \frac{\T R^2}{\sqrt{\lambda}}~.
\end{equation}
Here $l_{s}$ is the string length and $\lambda$ is the t’Hooft coupling parameter. It should be mentioned that for a pure AdS spacetime, the Weyl tensor square vanishes, which leads to the vanishing of the one-point function of the boundary operator $\mathcal{O}$. Although for an arbitrary black brane background, the Weyl tensor square is non-zero, this eventually acts as a source to the bulk scalar field $\phi$.  The boundary one-point function at leading order in $\alpha$ is given by a Gubser, Klebanov, Polyakov, Witten (GKPW) \cite{Gubser:1998bc,Witten:1998qj}-like bulk-boundary dictionary of the schematic form \cite{Grinberg:2020fdj,David:2022nfn}
\begin{equation}
    \langle \mathcal{O}(t,x)\rangle \propto \alpha \int dz dt^{\prime} d^{d-1}x^{\prime} \sqrt{-g}~\mathcal{G}(t,x;z,t^{\prime},x^{\prime})W^2(z,t^{\prime},x^{\prime})
\end{equation}
where $\mathcal{G}$ is the propagator between the boundary ($x,t$) to the bulk point ($x^{\prime},t^{\prime}$) and $z$ is the bulk radial coordinate. The above expression can be written in a simpler form, which reads
\begin{equation}\label{one point greens fn}
    \langle \mathcal{O}(0)\rangle \propto \alpha \int d^{d+1}x \sqrt{-g}~ \mathcal{G}(0,x)W^2(x)~.
\end{equation}
Now, for a large mass that is $m\T R\gg1$, the conformal dimension is approximately $\Delta\sim mR$. This approximation leads to the following expression for the propagator $\mathcal{G}$, which reads
\begin{equation}
    \mathcal{G}\sim e^{-ml}~.
\end{equation}
In the above expression, $l$ is the renormalised proper length between the boundary and a bulk point. In order to evaluate the integral, we will mainly focus on the Saddle point contribution. At the saddle point, the primary solution occurs at a location where the influence of the propagator \(\mathcal{G}\) is counterbalanced by the \(W^2\) term. However, this does not happen anywhere outside the Euclidean black hole background. One can analytically continue the integral in eq.\eqref{one point greens fn} to the region near the singularity. At the singularity, the Weyl tensor square ($W^2$) is divergent. Therefore, it can balance the pull from the propagator. The analytic continuation using the branch with $l_{hor}+i\tau$ leads to a decaying exponential in eq.\eqref{one point greens fn}. Here $l_{hor}$ is the distance from the boundary to the black hole horizon and $\tau$ is the radial coordinate inside the black hole. This kind of decaying exponential is expected in saddle-point evolution. Therefore, it leads to the following saddle-point equation, which reads
\begin{equation}
    \Bigg[-i m + \frac{\partial}{\partial\tau}\log W^2 (\tau)\Bigg]_{\tau=\mathcal{T}_s}=0~.
\end{equation}
Usually, for black hole spacetimes, the Weyl tensor square $W^2 \propto (\tau-\tau_s)^{-\beta}$ inside the black hole. Here, $\beta$ is some positive constant, and it depends on the spacetime. Using this value of $W^2$ in the above saddle-point equation, we obtain
\begin{align}
    &-i m + \frac{\beta}{(\mathcal{T}_s -\tau_s)}=0\nonumber\\
    &\implies\mathcal{T}_s -\tau_s=\frac{-i\beta}{m}~.
\end{align}
Thus, in the leading order, one can evaluate the integral in eq.\eqref{one point greens fn} as follows
\begin{equation}\label{one point function}
    \langle\mathcal{O}\rangle\propto\sqrt{g(\mathcal{T}_s)}W^{2}(\mathcal{T}_s)e^{-m l(\mathcal{T}_s)}\propto e^{-m l_{hor}-i m \mathcal{T}_s}~.
\end{equation}
\begin{figure}
    \centering
    \includegraphics[width=0.9\linewidth]{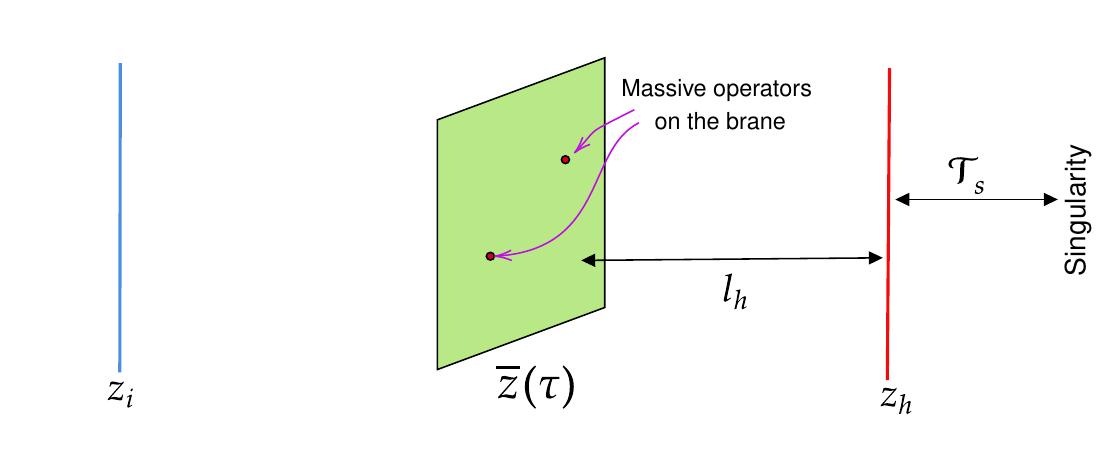}
    \caption{A schematic representation of the location of massive operators and the spacelike geodesic length from the brane to the horizon in the braneworld model and the timelike geodesic length from the horizon to the singularity. Here, $z_i$, $z_h$, and  $\z (\tau)$ are the initial brane position, horizon location, and the current brane position at a fixed cosmological time $\tau$, respectively. The massive operators with conformal dimension $\Delta$ are denoted by red dots on the brane.}
    \label{fig:braneworld}
\end{figure}
For heavy operators, we have already shown that the conformal dimension is approximately given by the operator mass $m$. This identification allows us to rewrite the above equation in the following form
\begin{equation}\label{one point function main}
    \langle\mathcal{O}\rangle \sim e^{-\Delta (l_{hor}+i  \mathcal{T}_s)}~.
\end{equation}
where $\Delta$ is the conformal dimension of the boundary operator $\mathcal{O}$ and the AdS radius $R$ has been set to one. The above expression will be helpful in deriving the thermal one-point functions for operators in the universe with various matter sources subsequently. In the context of the RS-II braneworld, the schematic diagram in Fig.\eqref{fig:braneworld} is useful while computing $l_h$ and $\mathcal{T}_s$ for a specific bulk AdS black brane geometry. 
\subsection{Single component universe}
Now that we have the expression for the thermal one-point function, we will proceed to evaluate the one-point function of an arbitrary operator in a single-component expanding universe.\\
We will start by calculating the regularised spacelike length from infinity to the horizon, which is given by 
\begin{equation}\label{lh}
    l_h=\int_{\z}^{z_h}\frac{dz}{z\sqrt{f(z)}}
\end{equation}
where $f(z)=1-\Big(\frac{z}{z_h}\Big)^{4-p}$.\\
Now, in order to evaluate the above integral, we will use the following binomial expansion
\begin{equation}\label{binomial}
    \frac{1}{\sqrt{1-x}}=\sum_{n=0}^{\infty}\frac{\Gamma(n+\frac{1}{2})}{\sqrt{\pi}\Gamma(n+1)}x^n~.
\end{equation}
Using the above-mentioned property, we can recast eq.\eqref{lh} in the following form. This reads
\begin{equation}
    l_h=\int_{\z}^{z_h}\frac{dz}{z}\sum_{n=0}^{\infty}\frac{\Gamma(n+\frac{1}{2})}{\sqrt{\pi}\Gamma(n+1)}\Big(\frac{z}{z_h}\Big)^{4n-np}~.
\end{equation}
Now, separating out the leading order term corresponding to $n=0$ from all of the other terms, we get
\begin{align}
    l_h&=\int_{\z}^{z_h}dz\Bigg[\frac{1}{z}+\frac{1}{z}\sum_{n=1}^{\infty}\frac{\Gamma(n+\frac{1}{2})}{\sqrt{\pi}\Gamma(n+1)}\Big(\frac{z}{z_h}\Big)^{4n-np}\Bigg]\nonumber\\
    &=\log\Big(\frac{z_h}{\z}\Big)+\sum_{n=1}^{\infty}\frac{\Gamma(n+\frac{1}{2})}{\sqrt{\pi}\Gamma(n+1)}\frac{1}{(4n-np)}+\dots~.
\end{align}
In the above equation, we have only written the leading-order terms in $\z$ as they will be useful later to obtain the leading-order time-dependent behavior of the thermal one-point function. After evaluating the sum in the second term, we get the final expression for $l_h$, which reads
\begin{equation}
    l_h=\log \Big(\frac{z_h}{\z}\Big)+\Bigg(\frac{\log 4}{4-p}\Bigg)~.
\end{equation}
The proper time the particle takes to reach the singularity from the horizon is given by the following integral
\begin{equation}
    \mathcal{T}_s=\int_{z_h}^{\infty}\frac{dz}{z\sqrt{\Big(\frac{z}{z_h}\Big)^{4-p}-1}}~.
\end{equation}
Again, using the binomial property from eq.\eqref{binomial} in the above equation, we obtain
\begin{equation}
    \mathcal{T}_s=\int_{z_h}^{\infty}\frac{dz}{z_h}\sum_{n=0}^{\infty}\frac{\Gamma(n+\frac{1}{2})}{\sqrt{\pi}\Gamma(n+1)}\Big(\frac{z_h}{z}\Big)^{4n-np+3-\frac{p}{2}}=\sum_{n=0}^{\infty}\frac{\Gamma(n+\frac{1}{2})}{\sqrt{\pi}\Gamma(n+1)}\frac{1}{(4n-np+2-\frac{p}{2})}~.
\end{equation}
Evaluating the above summation, we get the final expression for the proper time required to reach the singularity from the horizon
\begin{equation}
    \mathcal{T}_s=\frac{\pi}{(4-p)}~.
\end{equation}
Now with the expressions of $l_h$ and $\mathcal{T}_s$ in hand, we can use eq.\eqref{one point function} to obtain the thermal one-point function. Thus, the thermal one-point function for a universe with a single matter component is given by
\begin{equation}
    \langle \mathcal{O}\rangle _p \sim e^{-\Delta\log \Big(\frac{z_h}{\z}\Big)-\Delta\Big(\frac{\log 4}{4-p}\Big)}e^{\frac{-i\Delta\pi}{4-p}}~.
\end{equation}
The letter $p$ in the subscript indicates the $p$-brane gas configuration present in the bulk spacetime. This notation is useful for identifying the thermal one-point functions for different matter-dominated universes. The second exponential term just adds a phase factor to the overall result. This expression can be further simplified to obtain
\begin{equation}
    \langle \mathcal{O}\rangle _p \approx \Big(\frac{\z}{z_h}\Big)^{\Delta}2^{\frac{-2\Delta}{4-p}}\cross e^{\frac{-i\Delta\pi}{4-p}}~.
\end{equation}
In the above equation, we have only considered the leading-order terms in $\z$. For the RS-II braneworld model, the expansion of the universe is analogous to the time-dependent radial position of the brane. In an expanding universe, the physical observables must be time-dependent. Thus the obtained expression of the thermal one-point function must depend on the cosmological time $\tau$. Therefore, in order to include time dependence in our result, we will take the brane's radial position to be time-dependent \footnote{Claiming the brane's radial position is time dependent makes sense as the whole calculation is performed at a fixed cosmological time that is at a fixed brane position. This way of computing the time-dependent two-point function in the expanding universe is also used in \cite{Park:2024omh}.}.
\begin{equation}\label{opf in p}
    \langle \mathcal{O}\rangle _p(\tau) \approx \Big(\frac{\z (\tau)}{z_h}\Big)^{\Delta}2^{\frac{-2\Delta}{4-p}}\cross e^{\frac{-i\Delta\pi}{4-p}}~.
\end{equation}
The above expression of the thermal one-point function contains a complex term that is referred to as a phase factor. However, for the one-point function to be a physical quantity of the universe, we must ignore the phase factor or work with the modulus of the above expression of $\langle \mathcal{O}\rangle _p(\tau)$. From now on, we will ignore this complex phase factor. In this work, we are mainly focused on finding the leading-order time dependence of the one-point function of the universe in the presence of different matter sources. The phase factor indeed does not contain any term that can change the time dependence of the one-point function.
\subsubsection{Radiation dominated universe}
In the braneworld model, radiation is created by the back reaction of $0$-branes in the bulk. Therefore, in order to get the one-point function of a radiation-dominated universe, we will set $p=0$ in eq.\eqref{opf in p}. This gives
\begin{equation}
    \langle \mathcal{O}\rangle_{0}=\Big(\frac{\z}{z_h}\Big)^{\Delta}e^{-\Delta/2}~.
\end{equation}
Now we will proceed further to evaluate the early and late time behavior of this one-point function. Using the time-dependent brane positions for a radiation-dominated universe in the early times, we get
\begin{equation}
    \langle \mathcal{O}\rangle_{0}^{(early)}=\Big(\frac{z_i}{z_h}\Big)^{\Delta}\Bigg(1-\frac{\sqrt{2}z_i}{z_h}\sqrt{\tau}\Bigg)^{\Delta}e^{-\frac{\Delta}{2}}\approx \Big(\frac{z_i}{z_h}\Big)^{\Delta}\Bigg(1-\frac{\Delta\sqrt{2}z_i}{z_h}\sqrt{\tau}\Bigg)e^{-\frac{\Delta}{2}}~.
\end{equation}
\begin{figure}[htbp]
    \centering

    \begin{subfigure}[b]{0.49\textwidth}
        \centering
        \includegraphics[width=\textwidth]{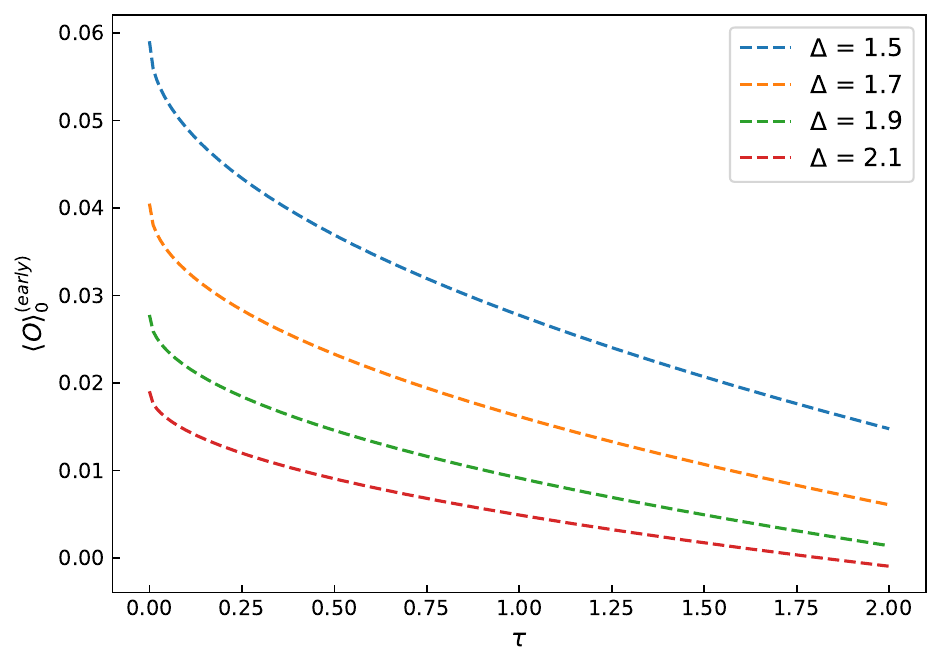}
        \caption{$\langle O \rangle_0^{(early)}$ vs $\tau$.}
        \label{fig:delta15}
    \end{subfigure}
    \hfill
    \begin{subfigure}[b]{0.49\textwidth}
        \centering
        \includegraphics[width=\textwidth]{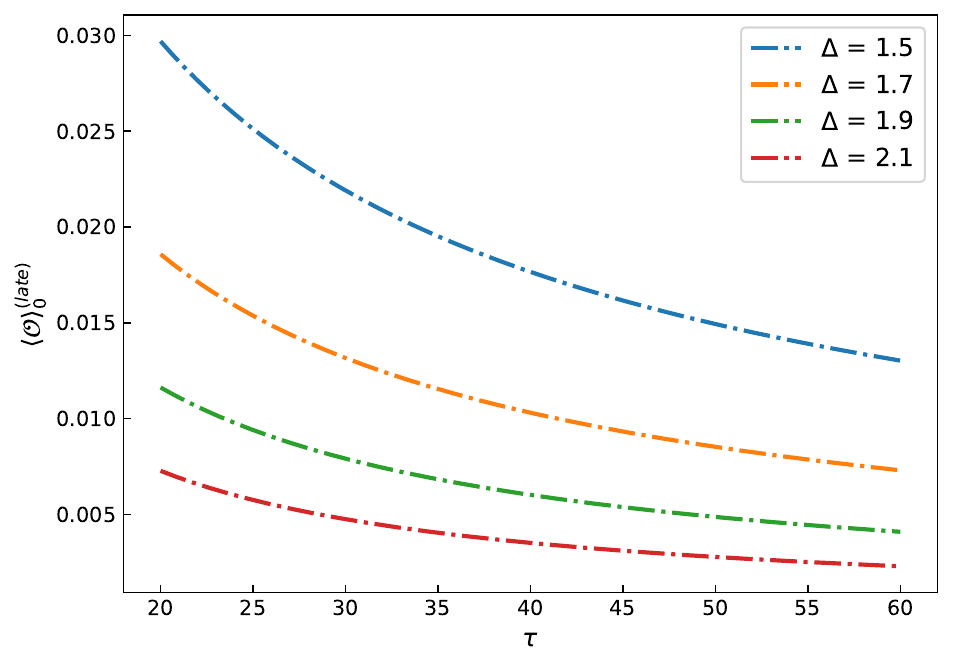}
        \caption{$\langle O \rangle_0^{(late)}$ vs $\tau$.}
        \label{fig:delta17}
    \end{subfigure}

    \caption{Early and late time behavior of $\langle O \rangle_0$ for different values of the conformal dimension $\Delta$. In both the figures, blue, orange, green and red lines are for $\Delta=1.5,~1.7,~1.9,~2.1$ respectively. The figure on the left panel is drawn for $z_i=0.5$ and $z_h=2$.}
    \label{fig:early_time_two}
\end{figure}
The above expression clearly shows that in the early time of a radiation-dominated universe, the thermal one-point function grows as $\sqrt{\tau}$ in the leading order of the cosmological time.
Similarly, using the late-time expression for the brane position for a radiation-dominated universe, one gets the following
\begin{equation}
    \langle \mathcal{O}\rangle_{0}^{(late)}=\frac{\tau^{-\frac{\Delta}{2}}}{2^{\frac{\Delta}{2}}}e^{-\frac{\Delta}{2}}~.
\end{equation}
In the late time, we can see that the one-point function changes as $\tau^{-\frac{\Delta}{2}}$. For operators with positive conformal dimension, the one-point function shows a falling nature with cosmic time $\tau$. \\
With the early and late time behavior of the one-point function for a purely radiation-dominated universe in hand, we will move forward to graphically represent our results. In Fig.(\eqref{fig:early_time_two}), we have plotted the early and late time expressions for the thermal one-point function in a radiation-dominated universe on the brane. The left and right panels show the early and late-time dynamics of $\langle \mathcal{O}\rangle_{0}$, respectively, for different values of conformal dimensions. In both figures, the blue, orange, green and red curves are for $\Delta=1.5,~1.7,~1.9,~2.1$ respectively. While obtaining these figures, we have chosen $z_i=0.5$ and $z_h=2$. Both of the graphs show the falling nature of the thermal one-point function, but at two different rates.
\subsubsection{Matter dominated universe}
As discussed earlier, bulk $1$-branes backreact on the brane on which the universe is situated. This backreaction results into matter in the FLRW universe. Thus, setting $p=1$ in eq.\eqref{opf in p} gives the thermal one-point function in a matter-dominated universe, which reads
\begin{equation}\label{opf matter}
    \langle \mathcal{O}\rangle_{1}=\Big(\frac{\z}{z_h}\Big)^{\Delta}e^{-2\Delta/3}~.
\end{equation}
The above expression is useful to compute the early and late-time behavior of the one-point function in a matter-dominated universe. Using the early-time brane dynamics equation in eq. , we get
\begin{equation}
    \langle \mathcal{O}\rangle_{1}^{(early)}=\Big(\frac{z_i}{z_h}\Big)^{\Delta}\Bigg(1-\Big(\frac{3}{2}\Big)^{\frac{2}{3}}\T\rc^{\frac{1}{3}}\tau^{\frac{2}{3}}z_i\Bigg)^{\Delta}e^{-2\Delta/3}\approx \Big(\frac{z_i}{z_h}\Big)^{\Delta}\Bigg(1-\Delta\Big(\frac{3}{2}\Big)^{\frac{2}{3}}\T\rc^{\frac{1}{3}}\tau^{\frac{2}{3}}z_i\Bigg)e^{-2\Delta/3}~.
\end{equation}
\begin{figure}[htbp]
    \centering

    \begin{subfigure}[b]{0.49\textwidth}
        \centering
        \includegraphics[width=\textwidth]{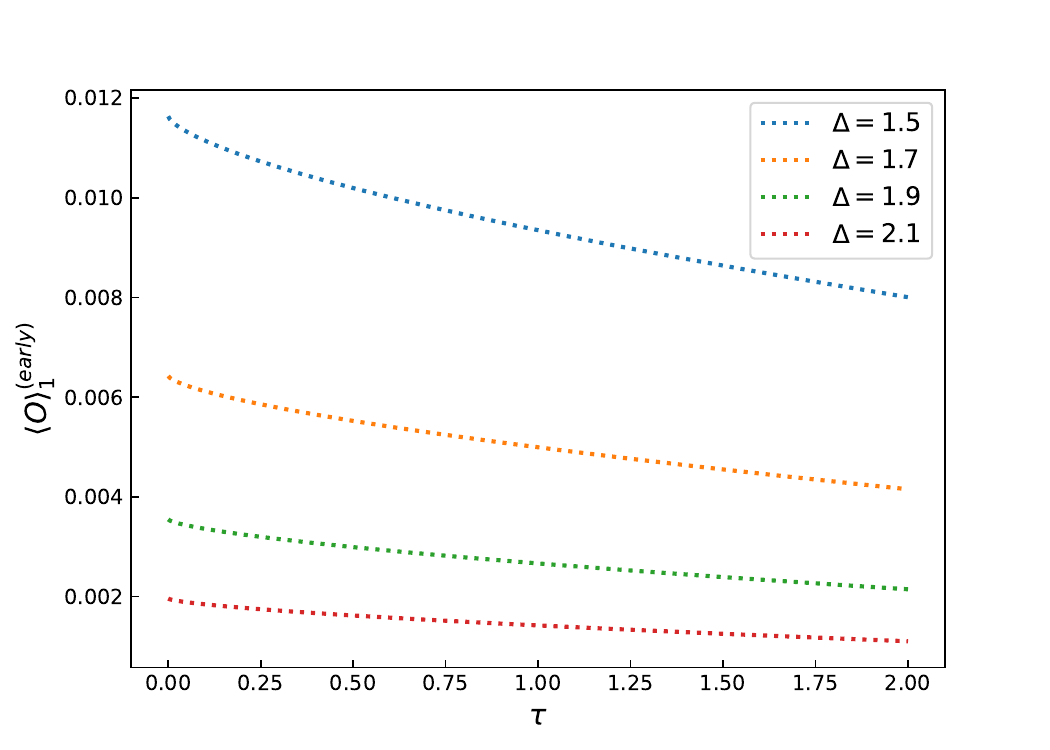}
        \caption{$\langle O \rangle_1^{(early)}$ vs $\tau$.}
        \label{fig:matter_early}
    \end{subfigure}
    \hfill
    \begin{subfigure}[b]{0.49\textwidth}
        \centering
        \includegraphics[width=\textwidth]{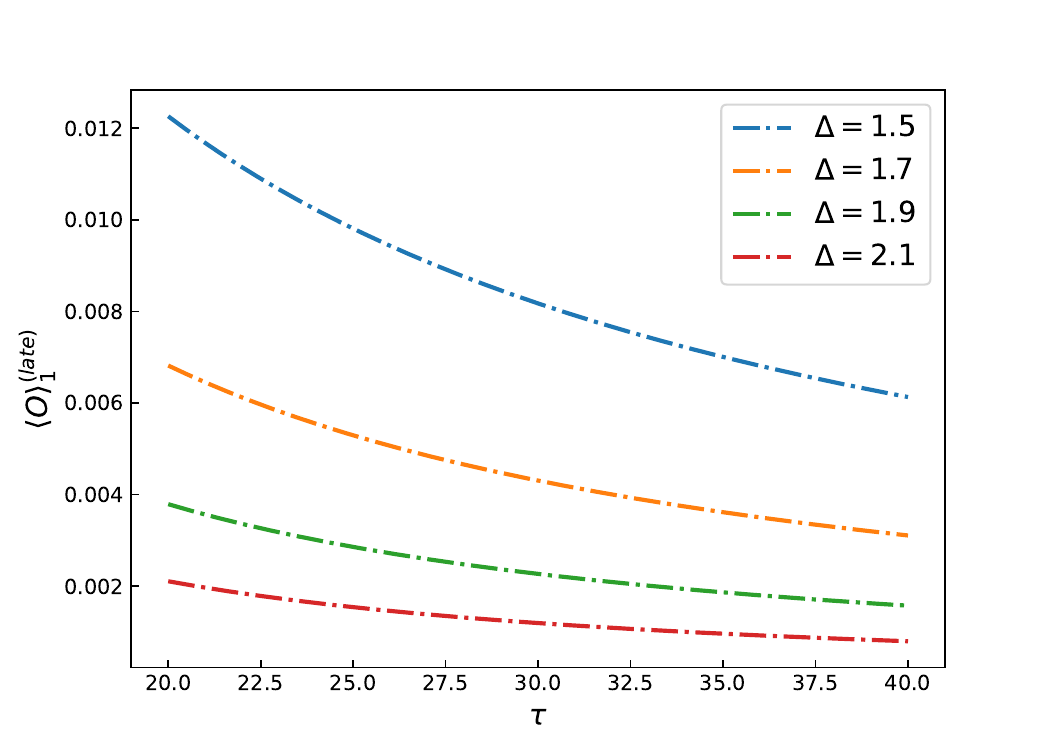}
        \caption{$\langle O \rangle_1^{(late)}$ vs $\tau$.}
        \label{fig:matter_late}
    \end{subfigure}

    \caption{Plot of the early and late-time behavior of \(\langle O \rangle_1\) for various conformal dimensions \(\Delta\). The blue, orange, green, and red lines correspond to \(\Delta = 1.5, 1.7, 1.9,\) and \(2.1\), respectively. The left panel is generated with initial position \(z_i=0.5\) and horizon position \(z_h=2\).}
    \label{fig:matter_early_late}
\end{figure}
In the early-time regime, the leading-order behavior of the one-point function in a matter-dominated universe scales as \(\tau^{2/3}\).
Similarly, using the late-time brane dynamics equation from eq. and substituting it into eq.\eqref{opf matter}, we get
\begin{equation}
    \langle \mathcal{O}\rangle_{1}^{(late)}=\Bigg(\frac{(\frac{2}{3})^{\frac{2}{3}}}{\T\rc^{\frac{1}{3}}z_h \tau^{\frac{2}{3}}}\Bigg)^{\Delta}e^{-2\Delta/3}~.
\end{equation}
Here $z_{h1}$ is the horizon radius of the black brane geometry in the presence of bulk one-branes. Thus, the late-time expression for the above one-point function can be written as 
\begin{equation}
    \langle \mathcal{O}\rangle_{1}^{(late)}=\Bigg(\frac{(\frac{2}{3})^{\frac{2}{3}}}{ \tau^{\frac{2}{3}}}\Bigg)^{\Delta}e^{-2\Delta/3}~.
\end{equation}
Thus the late-time behavior of \(\langle \mathcal{O} \rangle_{1}\) exhibits a power-law decay proportional to \(\tau^{-\frac{2\Delta}{3}}\). For better understanding, we have plotted the thermal one-point function with time in Fig.\eqref{fig:matter_early_late} for different values of conformal dimension. On the left side, we have shown the one-point function for early time, and on the right side, for late time. We have chosen initial brane position ($z_i$) and horizon position ($z_h$), $0.5$ and $2$, respectively. Four curves in different colours for each side of the figure related to $\Delta$ values $1.5,1.7,1.9$ and $2.1$. From the plot, we can see that the one-point function in the matter-dominated universe is falling with the cosmological time.
\subsubsection{Exotic matter dominated universe}
In the braneworld scenario, a string gas composed of 2-branes in the higher-dimensional bulk induces a form of exotic matter within the universe on the brane, characterised by an equation of state parameter $\omega = -\frac{1}{3}$. Therefore, using $p=2$ in eq.\eqref{opf in p} gives us the thermal one-point function of an exotic-matter-dominated universe. This reads
\begin{equation}
    \langle \mathcal{O}\rangle_{2}=\Big(\frac{\z}{z_h}\Big)^{\Delta}e^{-\Delta}~.
\end{equation}
We will now analyze the time evolution of this one-point function by considering the time-dependent positions of the branes within an exotic matter-dominated universe. In the early times, the one-point function for an exotic matter-dominated universe is given by
\begin{equation}
    \langle \mathcal{O}\rangle_{2}^{(early)}=\Big(\frac{z_i}{z_h}\Big)^{\Delta}\Big(1-\sqrt{\T\delta}z_i \tau\Big)^{\Delta}2^{-\Delta}\approx \Big(\frac{z_i}{z_h}\Big)^{\Delta}\Big(1-\Delta\sqrt{\T\delta}z_i \tau\Big)2^{-\Delta}~.
\end{equation}
\begin{figure}[htbp]
    \centering

    \begin{subfigure}[b]{0.49\textwidth}
        \centering
        \includegraphics[width=\textwidth]{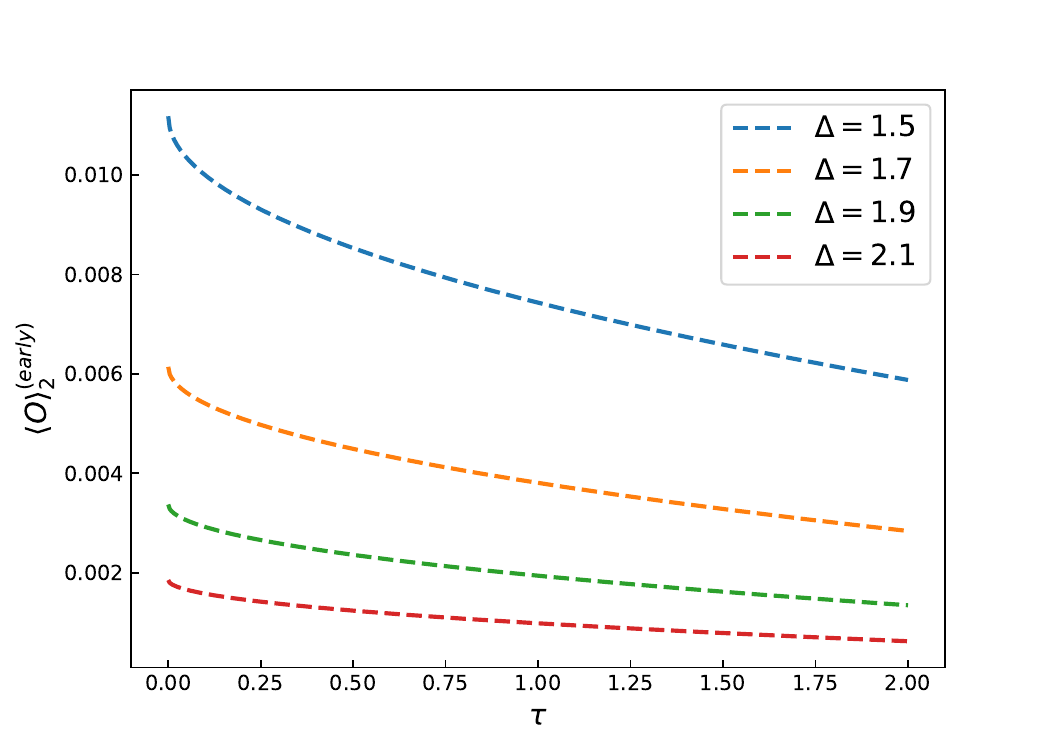}
        \caption{$\langle O \rangle_2^{(early)}$ vs $\tau$.}
        \label{fig:ex_early}
    \end{subfigure}
    \hfill
    \begin{subfigure}[b]{0.49\textwidth}
        \centering
        \includegraphics[width=\textwidth]{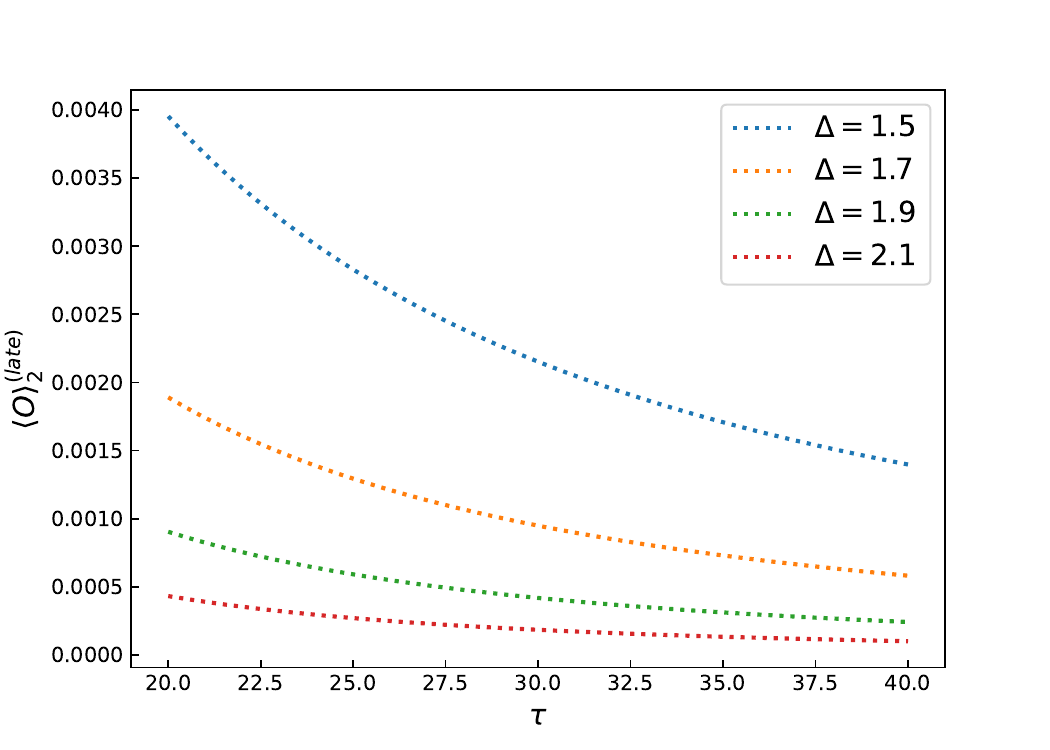}
        \caption{$\langle O \rangle_2^{(late)}$ vs $\tau$.}
        \label{fig:ex_late}
    \end{subfigure}

    \caption{The early and late-time behavior of \(\langle O \rangle_2\) for various conformal dimensions \(\Delta\). In both figures, the lines colored blue, orange, green, and red correspond to \(\Delta=1.5,\;1.7,\;1.9,\;2.1\), respectively. The left panel is generated with parameters \(z_i=0.5\) and \(z_h=2\).}
    \label{fig:exotic_matter_early_late}
\end{figure}
In an exotic matter-dominated universe, the early-time behavior of the one-point function shows growth proportional to $\tau$ in the leading order.
In the late times, the one-point function is given by
\begin{equation}
    \langle \mathcal{O}\rangle_{2}^{(late)}=\Bigg(\frac{1}{\sqrt{\T\delta}\tau z_h}\Bigg)^{\Delta}2^{-\Delta}~.
\end{equation}
Just like the other matter-dominated universes, here also the one-point function shows a fall in the late-time regime. Here, the one-point function decays as a function of the cosmological time, proportional to $\tau^{-\Delta}$. Similar to the previous cases of radiation and matter-dominated universes, here also we have shown the dynamics of the one-point function with time with different conformal dimensions $\Delta=1.5,1.7,1.9$ and $2.1$. The early and late time behavior is shown in the left and right panels, respectively. While plotting the curves, we have set the initial brane position and horizon to $0.5$ and $2$, respectively. Similar to the other two cases, here too the one-point function is falling with time, both in the early and late time, but with different rates \\
A few comments regarding the early and late time behavior of the one-point function in different matter-dominated universes (radiation, matter and exotic matter) are in order. At early times, when the brane just begins to move, the spacelike geodesic stretching from a massive operator on the brane to the horizon starts to shrink. Because the brane is moving relatively fast in this stage, the geodesic length decreases quickly. This rapid change is what gives rise to the polynomial decay observed in the thermal one-point function across different matter-dominated cosmologies.
As the system evolves, the brane gets very close to the horizon. The portion of the geodesic that lies inside the horizon nearly reaches a singularity, signalling the onset of a quasi-static phase. In this regime, the decay of the one-point function becomes much slower and can be treated as approximately static.
At late time, however, a different effect becomes important. As the brane approaches the horizon even more closely, the information associated with the massive operator is effectively absorbed by the black brane horizon. This can be interpreted as a form of information loss, and it leads to a power-law decay of the thermal one-point function in various matter-dominated universes. One might expect that as the brane slows down at late times, the spacelike geodesic length would decrease more gradually, leading to a slower decay of the one-point function. However, in this regime, the absorption of information by the horizon becomes the dominant effect, outweighing the slow change in the geodesic length $l_h$. As a result, the thermal one-point function actually decays more rapidly at late times in various matter-dominated universes.
In universes containing multiple components, similar behavior of the one-point function is observed as in universes with a single matter component. For universes with coexisting radiation and matter, the early-time behavior of the one-point function is dominated by radiation density, while at late times, matter density governs its dynamics. This pattern also holds for universes with coexisting radiation and exotic matter.
\subsection{Multi component universe}
Till now, we have only discussed about the one-point function of an operator in the universe with a single matter component. Although in a more realistic scenario, the universe has multiple coexisting matter sources. Therefore, it is necessary to obtain the expression for the one-point function in a multi-component universe. In a previous study \cite{10.1093/ptep/ptag115}, we have proposed a systematic way to study the entanglement properties in the braneworld scenario through a generalised solution of black brane geometry with $p$-brane gas. Here, we will also use the same lapse function as discussed in \cite{10.1093/ptep/ptag115} to compute the one-point function of an operator in the multi-component universe.
\subsubsection{Universe with coexisting radiation and matter}
For a universe with coexisting radiation and matter, the black brane lapse function in the braneworld model is given by
\begin{equation}
    f(z)=1-\T\rc z^3 -\m z^4~.
\end{equation}
Using this lapse function, we can compute the spacelike geodesic length from the brane up to the horizon, which reads
\begin{equation}
    l_h=\int_{\z}^{z_h}\frac{dz}{z\sqrt{1-\T\rc z^3 -\m z^4}}~.
\end{equation}
We can solve the above integral perturbatively. Considering $\T\rc$ and $\T\m$ to be small perturbation parameters, we can write
\begin{align}
    l_h&=\int_{\z}^{z_h}\frac{dz}{z}\Bigg[1+\frac{\T\rc}{2} z^3 +\frac{\T\m}{2} z^4\Bigg]+\mathcal{O}(\T\m^2)+\mathcal{O}(\T\rc^2)+\dots\nonumber\\
    &\approx \log\Big(\frac{z_h}{\z}\Big)+\frac{\T\rc}{6}(z_h^3 -\z^3)+\frac{\m}{8}(z_h^4 -\z^4)~.
\end{align}
In the last line of the above equation, we have neglected the quadratic and all the next higher-order terms of $\T\rc$ and $\T\m$.
Similarly the time-like geodesic extending from the horizon to the singularity can be computed through the following integral
\begin{equation}
    \mathcal{T}_s=\int^{\infty}_{z_h}\frac{dz}{z\sqrt{\T\rc z^3 +\m z^4-1}}~.
\end{equation}
After choosing the variable transformation such that $z=\frac{z_h}{u}$, the above integral becomes
\begin{equation}
    \mathcal{T}_s=\int^{1}_{0}\frac{u}{\sqrt{1-u^4 -\T\rc z_h^3 (1-u)}}du~.
\end{equation}
As $\T\rc$ is a small parameter, we can approximately write
\begin{align}\label{tau s integral}
    \mathcal{T}_s&=\int^{1}_{0}\frac{u}{\sqrt{1-u^4}}\Bigg[1+\frac{\T\rc z_h^3 (1-u)}{2(1-u^4)}\Bigg]du+\mathcal{O}(\T\rc^2)+\dots\nonumber\\
    &\approx I_1 +\frac{\T\rc z_h^3}{2}I_2~.
\end{align}
where
\begin{align}
    I_1 &=\int_{0}^{1}\frac{u}{\sqrt{1-u^4}}du=\frac{\pi}{4}\\
    I_2&=\int_{0}^{1}\frac{u (1-u)}{(1-u^4)^{\frac{3}{2}}}du=\frac{1}{4}\Bigg[B\Big(\frac{1}{2},-\frac{1}{2}\Big)-B\Big(\frac{3}{4},-\frac{1}{2}\Big)\Bigg]~.
\end{align}
Here $B(x,y)=\int_{0}^{1}dp~p^{x-1}(1-p)^{y-1}$ is the beta function. Therefore, substituting these expressions of $I_1$ and $I_2$ into eq.\eqref{tau s integral}, we finally obtain
\begin{equation}
    \mathcal{T}_s=\frac{\pi}{4}+\frac{\T\rc z_h^3}{8}\Bigg[B\Big(\frac{1}{2},-\frac{1}{2}\Big)-B\Big(\frac{3}{4},-\frac{1}{2}\Big)\Bigg]~.
\end{equation}
The above expression of $\mathcal{T}_s$ clearly suggests that the temporal geodesic length inside the horizon has a constant value and it does not depend on the time-dependent brane position. This indicates that $\mathcal{T}_s$ is independent of the cosmological time $\tau$. Hence $\mathcal{T}_s$ only introduces a constant phase factor to the one-point function in the multi-component scenario. Now, using these expressions of $l_h$ and $\mathcal{T}_s$ in eq.\eqref{one point function main}, we can finally calculate the one-point function of an operator in a universe with coexisting radiation and matter. Therefore the one-point function in this scenario is given by
\begin{equation}\label{opf rad+mat gen}
    \langle\mathcal{O}\rangle_{rad+mat}=e^{-\Delta\Bigg[\log\Big(\frac{z_h}{\z}\Big)+\frac{\T\rc}{6}(z_h^3 -\z^3)+\frac{\T\m}{8}(z_h^4 -\z^4)\Bigg]}\cross e^{-i\Delta \Bigg[\frac{\pi}{4}+\frac{\T\rc z_h^3}{8}\Bigg(B\Big(\frac{1}{2},-\frac{1}{2}\Big)-B\Big(\frac{3}{4},-\frac{1}{2}\Big)\Bigg)\Bigg]}~.
\end{equation}
The second exponential in the above equation only contributes a constant phase factor. If we take the one-point function $\langle\mathcal{O}\rangle_{rad+mat}$ to be a physical quantity, we must work with the modulus of the above equation or neglect the phase factor.  A little bit of algebraic simplification shows
\begin{equation}
    \langle\mathcal{O}\rangle_{rad+mat}=\Big(\frac{\z}{z_h}\Big)^{\Delta} e^{-\Delta\Bigg[\frac{\T\rc}{6}(z_h^3 -\z^3)+\frac{\m}{8}(z_h^4 -\z^4)\Bigg] }~.
\end{equation}
With this expression in hand, we will now move forward to calculate the leading-order early and late-time behavior of $\langle\mathcal{O}\rangle_{rad+mat}$. In early times, the brane dynamics of a universe with coexisting radiation and matter are governed by eq.\eqref{early_brane_rad+matt}. Using this expression of $\z (\tau)$ in the above expression of the one-point function, we get
\begin{equation}
    \langle\mathcal{O}\rangle_{rad+mat}^{(early)}\approx\Big(\frac{z_i}{z_h}\Big)^{\Delta} e^{-\Delta\Bigg[\frac{\T\rc}{6}(z_i^3 -z_h^3)+\frac{\m}{8}(z_i^4 -z_h^4)\Bigg] }\cross \Bigg[1-\Delta\sqrt{\m}\tau \Big(z_i^2 +\frac{z_i z_t}{2}\Big)\Bigg]~.
\end{equation}
In the above equation, we have only considered terms up to the leading order in $\T m$ and $\T \rc$. We can clearly see that in the early times, the one-point function grows as $\tau$ for a universe with coexisting radiation and matter. The time-dependent part of the above equation is multiplied by a factor $\sqrt{\T m}$. This indicates that at an early time, the dynamics of the one-point function are mainly controlled by the radiation density, and the sub-leading corrections consist of matter density. This observation is consistent with the thermal history of the universe \cite{WMAP:2010qai,WMAP:2010sfg,Planck:2014loa,Planck:2018vyg}, where radiation had a dominant contribution in the early times.\\
The late-time dynamics of the brane in this kind of universe is given by eq.\eqref{late_brane_rad+matt}. By the use of this expression of $\z (\tau)$ in eq.\eqref{opf rad+mat gen}, one gets the following late time behavior of $\langle\mathcal{O}\rangle_{rad+mat}$
\begin{align}
    \langle\mathcal{O}\rangle_{rad+mat}^{(late)}&=\Bigg(\frac{\Big(\frac{2}{3}\Big)^{\frac{2}{3}}\T\rc^{-\frac{1}{3}}}{z_h}\Bigg)^{\Delta}\tau^{-\frac{2\Delta}{3}}\Bigg[1+\frac{\Delta\Big(\frac{2}{3}\Big)^{\frac{2}{3}}\T\rc^{-\frac{1}{3}}\tau^{-\frac{2}{3}}}{z_t}\Bigg]e^{-\Delta\Bigg[\frac{\T\rc}{6}z_h^3 +\frac{\T\m}{8}z_h^4 \Bigg]}\nonumber\\&\cross e^{\Bigg[\frac{\Delta\T\rc \tau^{-2}\Big(\frac{2}{3}\Big)^{2}\T\rc^{-1}\tau^{-2}}{6}+\frac{\Delta\T m \Big(\frac{2}{3}\Big)^{\frac{8}{3}}\T\rc^{-\frac{4}{3}}\tau^{-\frac{8}{3}}}{8}\Bigg]}~.
\end{align}
The above late-time expression for the one-point function shows a power-law decay with respect to the cosmological time. This kind of similar behavior was also observed for single-component universes. In the late-time regime of a universe with co-existing radiation and matter, the one-point function changes as $\tau^{-\frac{2\Delta}{3}}$. The leading-order dependence mainly comes from the contribution from matter, and the sub-leading corrections occur from radiation. This observation is also consistent with the thermal history of the universe \cite{WMAP:2010qai,WMAP:2010sfg,Planck:2014loa,Planck:2018vyg}, which states that in the present time, that is, the late-time, the universe is dominated by matter. 
\subsubsection{Universe with coexisting radiation and exotic matter}
For a universe with coexisting radiation and exotic matter, the black brane lapse function in the braneworld model is given by
\begin{equation}
    f(z)=1-\T\delta z^2 - \m z^4~.
\end{equation}
Using this lapse function, we can compute the spacelike geodesic length from the brane up to the horizon, which reads
\begin{equation}
    l_h=\int_{\z}^{z_h}\frac{dz}{z\sqrt{1-\T\delta z^2 -\m z^4}}~.
\end{equation}
As discussed earlier, for a universe with coexisting radiation and matter, we can also solve the above integral perturbatively. By considering $\T\delta$ and $\m$ to be small perturbation parameters, we can write the following expression for $l_h$
\begin{align}
    l_h&=\int_{\z}^{z_h}\frac{dz}{z}\Bigg[1+\frac{\T\delta}{2} z^2 +\frac{\m}{2} z^4\Bigg]+\mathcal{O}(\T\delta^2)+\mathcal{O}(\m^2)+\dots\nonumber\\
    &\approx \log\Big(\frac{z_h}{\z}\Big)+\frac{\T\delta}{4}(z_h^2 -\z^2)+\frac{\m}{8}(z_h^4 -\z^4)~.
\end{align}
In the last line of the above equation, we have neglected the quadratic and all the next higher-order terms of $\T\delta$ and $\m$.
Similarly, the time-like geodesic extending from the horizon to the singularity can be computed through the following integral
\begin{equation}
    \mathcal{T}_s=\int^{\infty}_{z_h}\frac{dz}{z\sqrt{\T\delta z^2 +\m z^4-1}}~.
\end{equation}
In order to solve the above integral, we will again follow a similar procedure that was done for a universe with coexisting radiation and matter. After doing the variable transformation $z=\frac{z_h}{p}$, we can write
\begin{equation}
    \mathcal{T}_s=\int^{1}_{0}\frac{p}{\sqrt{1-p^4-\T\delta z_h^2 (1-p^2)}}dp~.
\end{equation}
Now performing a series expansion of the integrand, considering $\T\delta$ to be a small parameter, we get
\begin{equation}\label{tau s rad+ex}
    \mathcal{T}_s=\int^{1}_{0}\frac{p}{\sqrt{1-p^4}}\Bigg[1+\frac{\T\delta z_h^2 (1-p^2)}{2 (1-p^4)}\Bigg]=\T I_1 +\frac{\T\delta z_h^2}{2}\T I_2
\end{equation}
where
\begin{align}
    \T I_1 &=\int_{0}^{1}\frac{p}{\sqrt{1-p^4}}dp=\frac{\pi}{4}\label{tilde I1}\\
    \T I_2&=\int_{0}^{1}\frac{p}{\sqrt{1-p^4}(1+p^2)}dp=\frac{1}{2}\label{tilde I2}~.
\end{align}
We would like to mention that in the above equation of $\mathcal{T}_s$, we have only kept the terms up to $\mathcal{O}(\T\delta)$. Finally, using the expressions of $\T I_1$ and $\T I_2$ from eq(s).(\eqref{tilde I1},\eqref{tilde I2}) and substituting them back in to eq.\eqref{tau s rad+ex}, we obtain 
\begin{equation}
    \mathcal{T}_s=\frac{\pi}{4}+\frac{\T\delta z_h^2}{8}~.
\end{equation}
It can be clearly see that the time-like geodesic length from the horizon to the singularity is independent of the time-dependent brane position. Hence, for a universe with coexisting radiation and exotic matter, $\mathcal{T}_s$ does not introduce any time-dependence in the expression of the one-point function. It only gives a constant phase factor which only depends upon horizon position $z_h$ and the exotic matter density parameter $\T\delta$. 
Now, using these expressions of $l_h$ and $\mathcal{T}_s$ in eq.\eqref{one point function main}, we can finally calculate the one-point function of an operator in a universe with coexisting radiation and matter. Therefore, the one-point function in this scenario is given by
\begin{equation}\label{opf rad+ex gen}
    \langle\mathcal{O}\rangle_{rad+ex}=e^{-\Delta\Bigg[\log\Big(\frac{z_h}{\z}\Big)+\frac{\T\delta}{4}(z_h^2 -\z^2)+\frac{\T\m}{8}(z_h^4 -\z^4)\Bigg]}\cross e^{-i\Delta\Big(\frac{\pi}{4}+\frac{\T\delta z_h^2}{8}\Big)}~.
\end{equation}
Just like the universe with coexisting radiation and matter, the second exponential in the above expression for the one-point function of a universe with coexisting radiation and exotic matter only contributes a constant phase factor. If we take the one-point function $\langle\mathcal{O}\rangle_{rad+ex}$ to be a physical quantity, we must work with the modulus of the above equation or neglect the phase factor. Further simplification of the above equation leads to the following
\begin{equation}
    \langle\mathcal{O}\rangle_{rad+ex}=\Big(\frac{\z}{z_h}\Big)^{\Delta} e^{-\Delta\Bigg[\frac{\T\delta}{4}(z_h^2 -\z^2)+\frac{\m}{8}(z_h^4 -\z^4)\Bigg] }
\end{equation}
With this expression in hand, we will now move forward to calculate the leading-order early and late-time behavior of $\langle\mathcal{O}\rangle_{rad+ex}$. In early times, the brane dynamics of a universe with coexisting radiation and matter are governed by eq.\eqref{early brane rad+ex}. Using this expression of $\z (\tau)$ in the above expression of the one-point function, we get
\begin{equation}
    \langle\mathcal{O}\rangle_{rad+ex}^{(early)}\approx \Big(\frac{z_i}{z_h}\Big)^{\Delta} \Bigg(1-\Delta\sqrt{\m}\Big(z_i^2 -\frac{\T z_t^2}{2}\Big)\tau\Bigg) e^{-\Delta\Bigg[\frac{\T\delta}{4}(z_h^2 -z_i^2)+\frac{\m}{8}(z_h^4 -z_i^4)\Bigg] }~.
\end{equation}
We would like to mention that while obtaining the above equation, we have only considered terms up to the leading order in $\T m$ and $\T \delta$. It is evident that, during the early epoch, the one-point function changes proportionally to \(\tau\) in a universe containing both radiation and matter. The time-dependent component of the equation is scaled by a factor \(\sqrt{\T m}\), suggesting that the early-time behavior of the one-point function is predominantly governed by the radiation density. The contributions from matter density appear as sub-leading corrections. This observation aligns well with the thermal history of the universe \cite{WMAP:2010qai,WMAP:2010sfg,Planck:2014loa,Planck:2018vyg}, in which radiation played the dominant role during its early stages.\\
The late-time dynamics of the brane in this kind of universe is given by eq.\eqref{late brane rad+ex}. By the use of this expression of $\z (\tau)$ in eq.\eqref{opf rad+ex gen}, one gets the following late time behavior of $\langle\mathcal{O}\rangle_{rad+mat}$
\begin{align}
    \langle\mathcal{O}\rangle_{rad+ex}^{(late)}&= \frac{\tau^{-\Delta}}{z_h^{\Delta}\T \delta^{\frac{\Delta}{2}}}\Bigg[1-\frac{1}{\sqrt{\m}\tau}\Big(\frac{3}{8\T z_t^2}-\frac{1}{2 z_i^2}\Big)\Bigg]^{\Delta}e^{-\Delta\Bigg[\frac{\T\delta}{4}z_h^2 +\frac{\T\m}{8}z_h^4 \Bigg]}\nonumber\\
    &\cross e^{\frac{\Delta \T \delta}{4 \T \delta \tau^2}\Bigg[1-\frac{1}{\sqrt{\m}\tau}\Big(\frac{3}{8\T z_t^2}-\frac{1}{2 z_i^2}\Big)\Bigg]^2}~e^{\frac{\Delta\m}{8\T\delta^2 \tau^4}\Bigg[1-\frac{1}{\sqrt{\m}\tau}\Big(\frac{3}{8\T z_t^2}-\frac{1}{2 z_i^2}\Big)\Bigg]^{4}}~.
\end{align}
In the late-time regime, the cosmological time $\tau\to \infty$, therefore we can further approximate the above expression for the one-point function for a universe with co-existing radiation and exotic matter. Thus, expanding the expression for $\langle\mathcal{O}\rangle_{rad+ex}^{(late)}$ in powers of \( \frac{1}{\tau} \) and retaining only the leading-order terms in \( \m \) and \( \T \delta \) gives
\begin{equation}
    \langle\mathcal{O}\rangle_{rad+ex}^{(late)}\approx \frac{\tau^{-\Delta}}{z_h^{\Delta}\T \delta^{\frac{\Delta}{2}}}\Bigg[1-\frac{\Delta}{\sqrt{\m}\tau}\Big(\frac{3}{8\T z_t^2}-\frac{1}{2 z_i^2}\Big)\Bigg]e^{-\Delta\Bigg[\frac{\T\delta}{4}z_h^2 +\frac{\T\m}{8}z_h^4 \Bigg]}e^{\Delta\Bigg[\frac{1}{4\tau^2}+\frac{\m}{8\T \delta^2 \tau^4}\Bigg]}~.
\end{equation}
The late-time behavior of the one-point function exhibits a power-law decay with respect to the cosmological time. Similar behavior has been previously observed in single-component universes and universe with coexisting radiation and matter. In a universe containing both radiation and exotic matter, the one-point function scales as \(\tau^{-\Delta}\) in the late-time regime. The dominant contribution to this decay originates from exotic matter, with sub-leading corrections arising from radiation. This observation aligns with the established thermal history of the universe \cite{WMAP:2010qai,WMAP:2010sfg,Planck:2014loa,Planck:2018vyg}, which indicates that matter dominates the energy content at late times.
\section{Conclusion}\label{sec 5}
Now we will summarise our findings. We have started our discussion with a brief review on the Randall–Sundrum II braneworld model. In this model, our four-dimensional universe is situated on a brane that is embedded in a spacetime with one extra dimension. The time-dependent radial motion of the brane is analogous to the time-dependent expansion of the universe. It is also shown that in this model, we can identify the radial brane position as the scale factor of the universe. In order to realise various matter components on the braneworld cosmological model, different $p$-brane gas configurations are introduced in the bulk spacetime. We have shown how the back reaction of different $p$-brane gases creates single (radiation, matter and exotic matter) and multi-component (radiation-matter and radiation-exotic matter) universes on the brane that contains the four-dimensional universe. Then, using the five-dimensional black brane solution for the bulk spacetime in the presence of $p$-brane gas and the second Israel junction condition, we have obtained time-dependent radial brane positions for single and multi-component universes. The early and late time behavior of the brane positions is useful for obtaining various entanglement measures and correlations in single and multi-component universes.\\
Before moving to the holographic computation of the one-point function of massive operators in the braneworld model, we have given a quick review of the geodesic formula for the thermal one-point function in the black hole spacetime. The geodesic formula for the thermal one-point function comes from the fact that a non-minimal coupling of the scalar field with higher curvature terms like the Weyl tensor square or Gauss-Bonnet tensor gives rise to the thermal one-point function of a massive boundary operator. These higher derivative couplings act as a source to the non-vanishing thermal one-point function. We have used this geodesic formula along with the time-dependent brane positions to obtain the time-dependent one-point function for massive operators on the brane. At first, we have evaluated the early and late time behavior of the one-point function in a single-component universe for radiation, matter and exotic matter dominance, respectively. Then we have done a similar thing for a multi-component universe with coexisting radiation-matter and radiation-exotic matter. In the single-component universe in the early time, the one-point function shows a polynomial growth with respect to cosmological time $\tau$. For radiation, matter and exotic matter-dominated universes, the one-point function scales as $\tau^{\frac{1}{2}}$, $\tau^{\frac{2}{3}}$ and $\tau$ respectively in the early time. Although in the late time, the one-point correlation function falls off with different powers of $\tau$ for different matter-dominated universes. In the late time regime, it decays as $\tau^{-\frac{\Delta}{2}}$, $\tau^{-\frac{2\Delta}{3}}$ and $\tau^{-\Delta}$ respectively for radiation, matter and exotic matter dominated universes. At early times, when the brane has just started to move, the spacelike geodesic connecting a massive operator on the brane to the horizon begins to shrink. Since the brane is moving relatively quickly in this phase, the geodesic length decreases rapidly. This behavior naturally leads to the polynomial decay observed in the thermal one-point function across different matter-dominated cosmologies.
As the system evolves, the brane approaches the horizon, and the segment of the geodesic lying inside the horizon comes very close to a singularity. This marks the onset of a quasi-static regime, where the dynamics slow down, and the decay of the one-point function becomes much more gradual, effectively behaving as if it were static.
At late times, however, another effect takes over. As the brane moves even closer to the horizon, the information associated with the massive operator is effectively absorbed by the black brane horizon. This process, which can be viewed as a form of information loss, drives a power-law decay of the thermal one-point function in matter-dominated universes. Although one might expect the slowing motion of the brane to result in a slower decay, since the geodesic length $l_h$ is shrinking more gradually, the absorption of information by the horizon becomes the dominant factor. Consequently, the decay of the one-point function is actually faster at late times. In universes with multiple components, we observe a similar behavior of the one-point function as seen in universes with a single matter component. We have also graphically represented the variation of the thermal one-point function for various matter-dominated universes with respect to the cosmological time. In the early and late time regions, the one-point functions are plotted for four different values of the conformal dimension, that is, for $\Delta=1.5,~1.7,~1.9,~2.1$. These $\Delta$ values are chosen such that the BF bound is satisfied and the bulk scalar fields form a stable configuration. The graphical depiction of the one-point function also shows that, at both early and late times, larger conformal dimensions correspond to slower decay rates of the thermal one-point function.  For a universe with coexisting radiation and matter, we have seen that the early-time behavior of the one-point function is dominated by the radiation density and the late-time behavior by the matter density. Similarly, for a universe with co-existing radiation and exotic matter, the early-time dynamics of the thermal one-point function are dominated by the radiation density of the universe. In the late time, the one-point function has a dominant contribution from the exotic matter. These observations on the one-point function for both the multi-component universes (radiation-matter and radiation-exotic matter) are consistent with the thermal history of the universe \cite{WMAP:2010qai,WMAP:2010sfg,Planck:2014loa,Planck:2018vyg}. As a potential future direction for this research, one could incorporate higher curvature corrections or anisotropies into the braneworld model framework and conduct an initial investigation into the behavior of one- and two-point correlation functions. Additionally, exploring the Karach-Randall braneworld model as an alternative to the Randall-Sundrum model presents another promising direction for future study.
\section*{Acknowledgment}
SP thanks SNBNCBS for the Senior Research Fellowship. GG extends his heartfelt gratitude to CSIR, Govt. of India, for the Junior Research Fellowship.
\subsection*{Data Availability Statement}
This article has no associated data, or the data will not be deposited.
\subsection*{Code Availability Statement} 
This article has no associated code, or the code will not be deposited.
\bibliographystyle{JHEP.bst}
\bibliography{ref.bib}	
\end{document}